\documentclass[aps,prl,twocolumn,amsmath,a4paper,showpacs]{revtex4}
\usepackage[utf8]{inputenc} \usepackage{graphicx} \usepackage{units}

\begin{document}

\title{Versatile Digital GHz Phase Lock for External Cavity Diode
  Lasers}

\author{J\"urgen Appel} \email{jappel@nbi.dk} \affiliation{Niels Bohr
  Institute, Copenhagen University, 2100 K{\o}benhavn, Denmark}

\author{Andrew MacRae} \email{amacrae@qis.ucalgary.ca}

\author{A. I. Lvovsky} \email{lvov@ucalgary.ca} \affiliation{Institute
  for Quantum Information Science, University of Calgary, Alberta T2N
  1N4, Canada }

\date{\today}


\begin{abstract} We present a versatile, inexpensive and simple
  optical phase lock for applications in atomic physics experiments.
  Thanks to all-digital phase detection and implementation of beat
  frequency pre-scaling, the apparatus requires no microwave-range
  reference input, and permits phase locking at frequency differences
  ranging from sub-MHz to 7 GHz (and with minor extension, to 12 GHz).
  The locking range thus covers ground state hyperfine splittings of
  all alkali metals, which makes this system a universal tool for many
  experiments on coherent interaction between light and atoms.
\end{abstract}

\maketitle
\section{Introduction}
Many experiments on coherent interaction of light with matter require
two or more separate light fields whose frequency difference is
precisely maintained at a particular value.  If the required frequency
difference is relatively small, the mutually coherent fields can be
produced by splitting the light from a single source and shifting the
frequency by acousto-optic or electro-optic modulation. But if the
frequencies of the required fields differ by more than a few hundred
MHz, frequency modulation techniques become impractical and it is
often preferable to use two separate sources.  In this case, long-term
phase coherence between these sources can be achieved using an optical
phase locked loop (OPLL). In such an OPLL, the rate at which the
relative phase between a master laser and a slave laser changes is
locked to a fixed value $f_\text{beat}$ called the beat frequency.

A number of OPLL designs have been implemented
\cite{telle93:_stabil,Prevedelli,cacciapuoti,Marino,Gouet2008_gravimeter}.
However, most of the existing schemes are designed with a specific
setup in mind and require expertise in microwave frequency
electronics. Furthermore, these schemes require expensive equipment
for the generation of stable microwave-range reference signals.  In
addition, the range of lockable beat frequencies is often restricted
to less than an octave and a change in the experimental configuration
requires major modifications in the hardware.

In this work, we present a versatile OPLL for use with external cavity
diode lasers (ECDLs) which is simple to construct and is made from
inexpensive off-the-shelf components. Recent advances in digital phase
detection, associated with the development of wireless communications,
allow us to develop a robust design which is insensitive to beat
signal amplitudes over a wide range and avoids the frequency range
limitations of passive microwave mixers. Since the beat frequency is
divided down digitally, a versatile and inexpensive low frequency
reference may be used. The high-frequency part of the printed circuit
board layout is kept compact, so constructing the device does not
require high frequency electronic measuring equipment and expertise.

In order to test the performance of the OPLL, the spectrum of the beat
signal was measured around $f_\text{beat}=\unit[6.9]{GHz}$ and the
phase variance $\langle(\Delta \phi)^2\rangle$ is found to be less
than $\unit[0.08]{rad^2}$.  Two lasers, each locked to the same source, were
interfered and steady fringes were observed.

\section{Construction of the Device}

\begin{figure}
  \includegraphics[width=80mm]{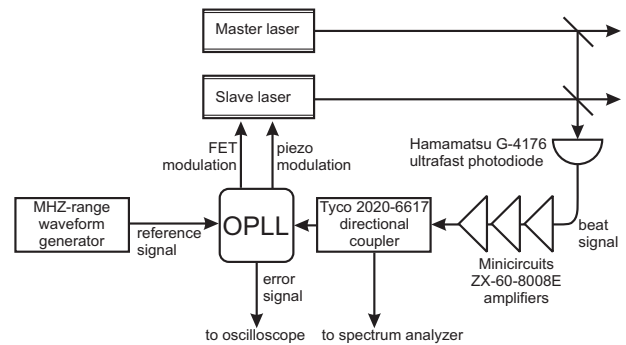}
  \caption{Experimental setup implementing our phase lock.}\label{fig:OpticalSetup}
\end{figure}

The basic setup for our OPLL is displayed in
figure~\ref{fig:OpticalSetup}. Beams from the master and slave lasers
are mixed at a beam splitter and a total optical power of
$\unit[0.5 -2 ]{mW}$ is focused onto a fast photodetector.  The generated
beat signal is amplified up to $\approx \unit[0]{dBm}$.
A~\mbox{\unit[$-$10]{dB}} directional coupler splits the beat
frequency between a spectrum analyzer and the OPLL circuitry.  Based
on a reference signal received from a function generator, the OPLL
produces the error signal for the slave laser, closing the feedback
loop.

The coupler and the spectrum analyzer are not required for daily
operation but prove to be useful to monitor lock performance when
initially setting up the system. 

Since the instantaneous bandwidth of a typical ECDL may span as much
as a few hundreds of kilohertz, the loop must be made as fast as
possible in order to correct for this noise. The loop must as well be
able to correct for the low frequency noise and drifts due to
mechanical vibrations. To achieve the required frequency range, dual
feedback was employed: by modulating the external cavity length with a
piezo-electric transducer, and by direct modulation of the injection
current. The bandwidth of the piezo modulation is limited to a few kHz
by mechanical resonances, but it allows significant correction to the
laser frequency. On the other hand, the current injection modulation
is fast (up to a few MHz), but the frequency corrections are
limited by modehops.

\begin{figure*}
  \includegraphics[width = \textwidth]{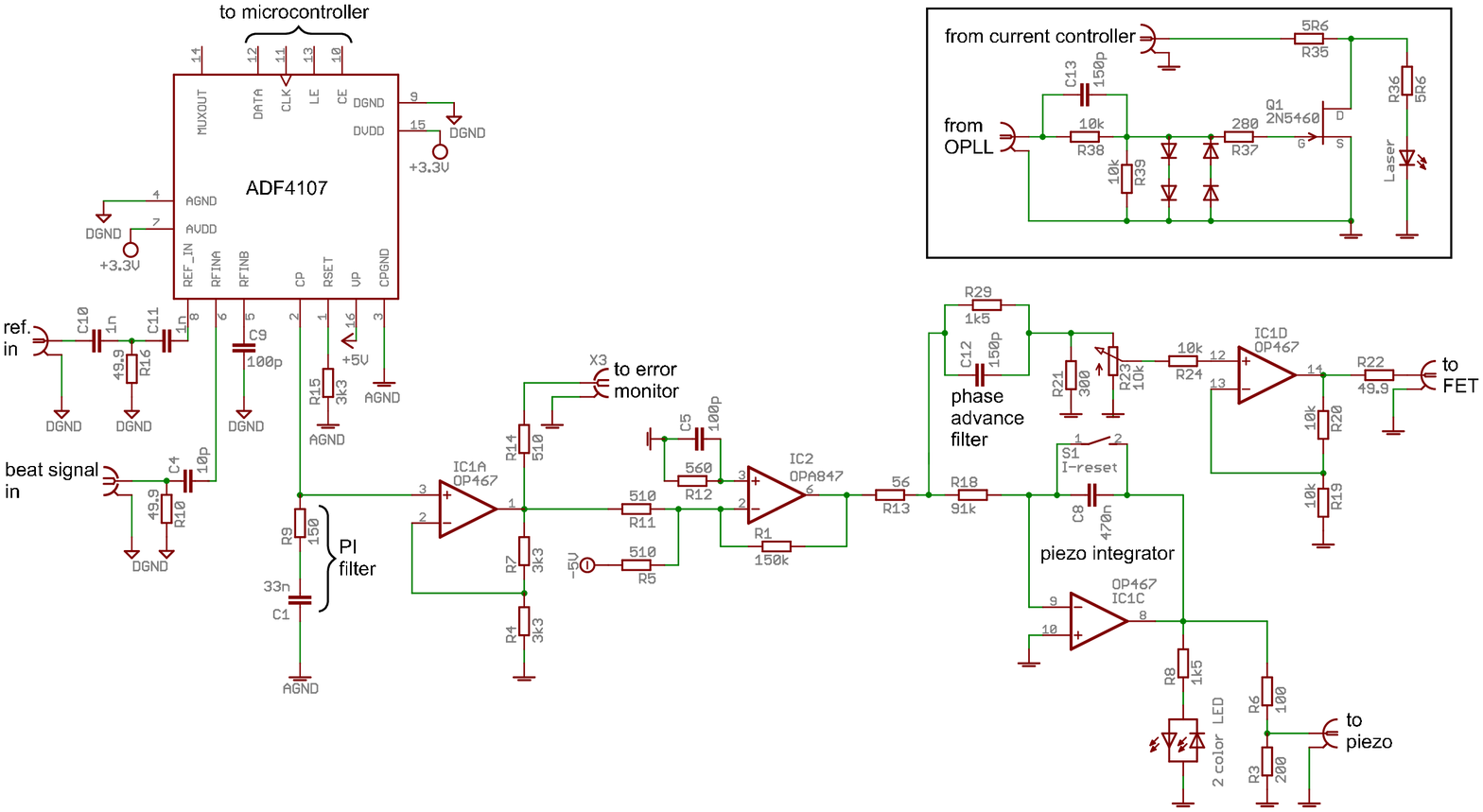}
  \caption{Circuit diagram of the OPLL controller.  Inset: Laser diode
    current modulator} \label{fig:circuit}
\end{figure*}

The electric schematic of our OPLL is shown in
figure~\ref{fig:circuit}. The reference and beat signals are sent into
the digital phase-frequency-discriminator chip (\emph{Analog devices
  ADF4107}).  This chip divides the reference signal by a factor $R$
and the beat signal by a factor $N$ digitally and then compares
frequency and phase of both divided signals with a dual flip-flop
circuit. The ADF4107 is interfaced with a micro-controller that
programs the divider quotients $R$ and $N$ on start-up and allows for
easy change of these values. A similar circuit based on the ADF4007
that does not require programming but is less versatile has been
constructed and shows similar performance.

The $N$ counter can be programmed to any value between 24 and about
$5\times 10^5$; the range of permitted values of $R$ can take values
between 1 and 16,383. The phase lock can thus be implemented for the
beat signal of any frequency up to the maximum permitted by the
ADF4107 (7 GHz), with the reference produced by a generic MHz-range
signal generator. When needed, the frequency range can be extended by
preceding the OPLL beat signal input with a pre-divider
(\textit{Hittite HMC364S8G}).  The lock stability generally improves
with smaller values of $N/R$.  It is particularly important that the
divided reference frequency significantly exceed the required loop
bandwidth. 

If the phase difference of the beat and reference signals is small,
this phase-frequency-discriminator circuit produces a feedback current
proportional to this quantity.  Otherwise the error signal polarity
corresponds to the sign of the frequency difference between the two
signals \cite{curtin1999:_PLL}.  In this way, the capture range of the
phase lock circuit is limited only by the mode hop free tuning range
of the slave laser.

The feedback current enters a proportional-integral (PI) filter formed
by C1, R9 and is amplified in two stages. Since the ADF4107's output
voltage is restricted to a range of \unit[0--5]{V}, a gain-two
preamplifier (IC1A) converts the voltage on the PI Filter into a low
impedance \unit[0--10]{V} output. A negative \unit[5]{V} bias is added
so that the output of the second amplification stage (IC2) with gain
300 is symmetric around zero when locked. This bias voltage has been
strongly low pass filtered to to avoid introducing noise to the error
signal (using IC1B, not depicted in Fig. 2). Finally the output of IC2
is split into the fast and slow feedback paths.

A characteristic challenge in constructing a feedback circuit for a
diode laser is due to the shape of its transfer function's phase. At
low modulation frequencies, a change in the diode injection current
affects the lasing frequency mainly because of the modulation of the
recombination area's temperature. At high modulation frequencies, the
lasing frequency is affected due to current induced charge density
modulations which effect the refractive index of the gain medium.
Unfortunately these two mechanisms oppose each other, which leads to a
phase shift of $\unit[180]{^\circ}$ at modulation frequencies
typically between \unit[1--10]{MHz} \cite{kobayashi82:_direc_algaas}.
In order to partly compensate for this effect, a phase-advance loop filter is
used in the fast feedback path, followed by a buffer stage (IC1D) with
an adjustable gain to drive the laser diode current modulator.

The current feedback is implemented as depicted in the inset in
Fig.~\ref{fig:circuit} and is based on the FET modulation circuit for
Toptica DL100 lasers. The gate voltage of a N-junction field effect
transistor is clamped by protection diodes.  C13 approximately
compensates the low-pass formed by the diodes' and the J-FET's
capacitance.  The input voltage causes the J-FET to bypass a fraction
of the diode's supply current. This way the diode current never
exceeds that provided by the current controller, so the expensive
laser diode is protected.

We now describe the slow feedback path. The signal part is integrated
by IC1C and R18, C8 to control the length of the external cavity. The
integrator ensures that in the locked condition the output of stage
IC2 is zero on average, so that the current modulation is free of DC
components which may saturate the amplifier stages and drive the laser
closer towards a mode-hop. A bipolar LED conveniently displays the
integrator overrun, indicating the need for operator intervention.

The transfer function of the integrating arm controlling the piezo is
not critical to the lock performance. In fact, we choose the gain in
this path so high that, in the absence of the current feedback loop,
oscillations about the target frequency occur.  Once the fast current
feedback is engaged it easily counters this effect and renders the
loop stable as a whole.

Special care has been taken in order to obtain good noise performance.
Since digital switching noise can have a detrimental effect on the
operation, the analog and digital sections of the circuit each have
their own voltage regulators and are located on separate ground
planes.

This OPLL has been put in to operation in a variety of setups
including locking a commercial Toptica DL-100 diode laser to a
Coherent MBR Ti:Sapphire laser, a self made diode laser with an
Eagleyard laser diode to the Ti:Sapphire laser, and two self-made
diode lasers with Sharp and SDL diodes to each other.

\section{Characterization of Lock Performance}

An important parameter for measuring the performance of a PLL is the
mean-square phase error $\langle\Delta \phi^2\rangle$. This error can
be determined by measuring the power fraction of the carrier in the
electronic spectrum $P(\nu)$ of the beat signal \cite{Prevedelli}:
\begin{equation}\label{phasenoise}
  \exp(-\langle\Delta \phi^2\rangle) =\frac{
    P_\text{carrier}}{{\int}^{\infty}_{-\infty}P(\nu)d\nu}.
\end{equation}
To that end we set up the lock circuit to operate at the frequency
around 6.9 GHz and recorded the RF power spectrum in a 10 MHz
frequency span around this frequency with a Hewlett-Packard E4405B
spectrum analyzer. The spectrum analyzer resolution bandwidth was set
to 3 kHz, video bandwidth to 30 kHz, and the detector type to
\emph{average} \cite{appnote150}. During a 100-s scan, the analyzer
acquired a $n=8192$ point data set. In this way, the resolution
bandwidth is significantly larger than the frequency range associated
with each data point (1.22 kHz).

\begin{figure}
  \includegraphics[width = \columnwidth]{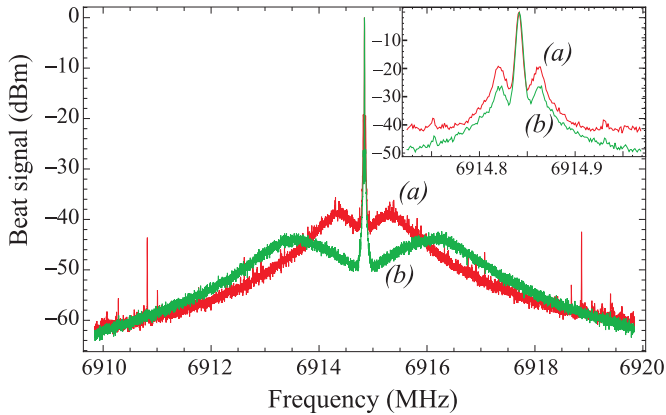}
  \caption{Spectral noise density of the RF signal produced by
    interference of two phase locked lasers (Resolution
    bandwidth=\unit[3]{kHz}). Curve (a) is associated with a 18 MHz
    reference signal produced by an Agilent 33220A reference
    generator; curve (b) with a 216 MHz home-made generator. The
    inset shows a blowout of the central portion of the
    plot.}\label{fig:BeatSpectrum}
\end{figure}

The beat spectrum is shown in Fig.~\ref{fig:BeatSpectrum}. Remarkably,
the width of the central peak, corresponding to the carrier signal,
could not be resolved even with the lowest resolution bandwidth
setting (10 Hz) of the spectrum analyzer.

The carrier power fraction is calculated from the acquired data set
$P_i$ (where $1\le i \le n$) as follows. The power in the carrier
$P_\text{carrier}$ is the direct reading of the spectrum analyzer at
the beat signal frequency. The surrounding noise power density
$P(\nu)$ can be determined from $P_i$ by applying the following
corrections \cite{spect_analy_measur_noise_an}. First, we add
\unit[2.51]{dB} to each $P_i$ to compensate the error associated with
the logarithmic scale envelope detection and subtract \unit[0.52]{dB}
to correct for the shape of the resolution bandwidth filter. Finally,
by subtracting $\unit[10 \log_{10}(3000)]{dB}$, we normalize the noise
power density to the 1-Hz bandwidth.

\begin{table}
  \centering
  \begin{tabular}{p{50mm}|l|l|l}
    Reference & $N$ & $R$ &  $(\Delta \phi)^2$ \\ \hline  \hline
    Agilent 33220A waveform generator at \unit[18]{MHz}  & 384 & 1    & $\unit[0.19]{rad^2}$ \\ \hline
    Agilent 33250A waveform generator at \unit[72]{MHz}  & 96 & 1     & $\unit[0.33]{rad^2}$\\ \hline
    Mini-Circuits JTOS 400, phase locked to \unit[216]{MHz} &  96 & 3   & $\unit[0.08]{rad^2}$ 
  \end{tabular}
  \caption{Phase noise}
  \label{tab:CarrierPower}
\end{table}

We performed the experiment with three reference oscillators, as
summarized in Table~\ref{tab:CarrierPower}. The best results were
obtained with a home-made generator consisting of a Mini-Circuits JTOS
400 voltage-controlled oscillator locked to an Agilent 33220A waveform
generator using an additional, narrowband ($\sim$1 kHz) phase-lock
circuit. The noise reduction associated with this generator is due to
a low multiplication factor $N/R$. Note that in spite of a lower $N$,
the lock with the Agilent 33250A reference generator (80 MHz) produces
more noise than Agilent 33220A (20 MHz). We believe this to be due to
the intrinsic phase noise of the reference signal, which is about
\unit[-115]{dBc/Hz} for 33220A and \unit[-90]{dBc/Hz} for 33250A.
These results emphasize the need for a low phase-noise reference
oscillator with large $N/R$. Additionally a high phase detection
frequency $f_\text{beat}/N$ that is significantly bigger than the loop
bandwidth is imperative to reduce the loop delay.

Ultimately even with a high quality reference oscillator the noise
performance achievable with the direct digital phase locking approach
presented in this work will be limited by the electronic noise floor
of the phase-frequency-discriminator circuit ADF4107, which is given by 
$\unit[-219]{dBc/Hz} \times N \times f_\text{beat}/\unit{Hz}$. Significantly lower
values can be obtained circumventing the frequency division by
multiplying up an ultra low noise frequency reference into the
microwave regime and then using that to mix down the laser beat note
signal. This however requires a much more complex and less versatile
setup\cite{Gouet2008_gravimeter}.

\begin{figure}
  \includegraphics[width = \columnwidth]{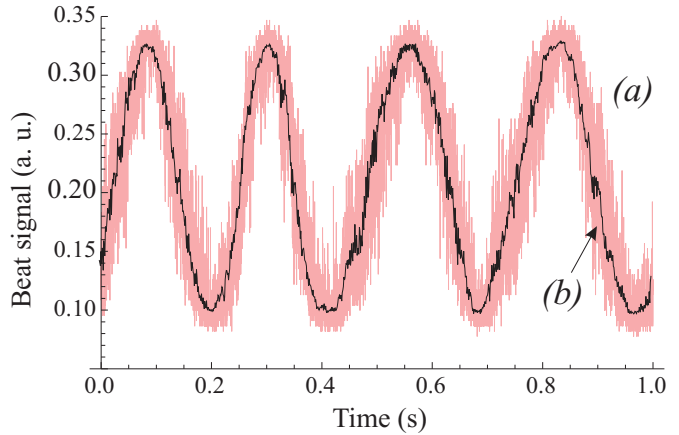}
  \caption{3.84-Hz beat signals between two slave lasers locked to the
    same master lasers, each with its own OPLL. Curve (a) represents a
    set of 1,000,000 data points; curve (b) the same set, smoothened
    by computing averages of 100-element bins.}
  \label{fig:beat}
\end{figure}

To illustrate the capabilities of our circuit, we have observed
interference between two separate slave lasers, phase locked to the
same master laser at 6.9 GHz and with a relative frequency difference
of 3.84 Hz. The interference signal between the two slave lasers was
measured with a fast photodiode and recorded with a digital
oscilloscope. The result of this measurement is displayed in
Fig.~\ref{fig:beat} and exhibits clearly discernible interference
fringes.

Further insight into the noise behavior of the phase lock is provided
by the modified Allan variance, which quantifies the average drift of
the oscillator frequency over time $\tau$
\cite{allan81:_modallan,rutmann91:_charac}. Using the $\unit[10]{MHz}$
output of a digital oscilloscope as reference we phase locked two
diode lasers to a common master laser at
$f_\text{beat}=\nu_0=\unit[4]{GHz}$. To measure the Allan variance, we
recorded four data sets of the interference signal between the slave
lasers, with sampling rates ranging between $4\times 10^9$ and $5\times
10^4$ s$^{-1}$. For short acquisition times, via a slow (bandwidth
$<\unit[1]{Hz}$) feedback loop, a piezoelectric transducer varied the
optical path lengths so that the interference pattern was held at
$50\%$ of its fringe. For acquisition times over \unit[0.5]{s}, the
piezo signal was disengaged and the measurement was started manually
after the pattern drifted to the $50\%$ fringe position.

\begin{figure}
  \includegraphics[width = \columnwidth]{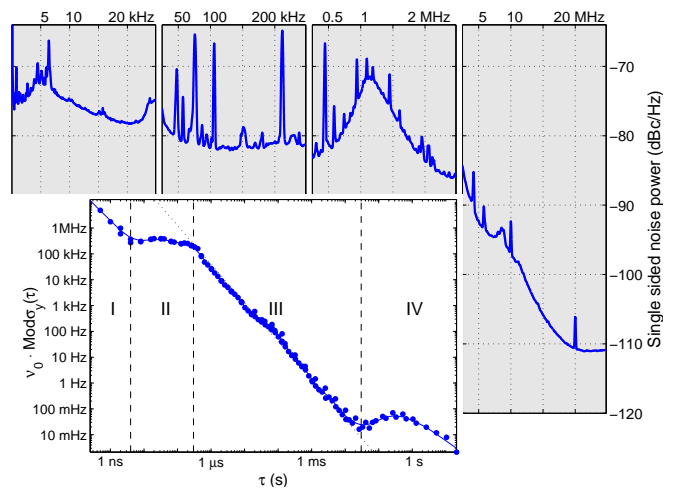}
  \caption{\textbf{Background:} Power spectrum of the interference
    signal of two lasers phase locked to a common master.
    \textbf{Inset:} Modified Allan deviation based on the same signal.
    Roman numerals indicate regions of different dominating noise
    types, see text.}
  \label{fig:MVAR}
\end{figure}

As depicted in the inset of Fig.~\ref{fig:MVAR}, the modified Allan
deviation (root Allan variance) decreases approximately as
$\tau^{-3/2}$ over a wide range of integration times between $\sim
300$ ns and $\sim 30$ ms (region III, black dotted line), which
corresponds to uncorrelated white phase noise
\cite{rutmann91:_charac}. Shorter $\tau$'s correspond to frequencies
outside the free-running linewidth of the diode laser, so the Allan
variance levels off ($1/f$ frequency noise, region II). At even
shorter integration times (below $\sim 10$ ns, region I), the
electronic noise of the photodetector comes into play, leading, again,
to a behavior similar to white phase noise. At very large $\tau$'s
(region IV), the phase measurement is disturbed by optical path length
fluctuations (vibrations, air flow), which show up as frequency and
phase drifts.

The power spectrum of the interference signal of the two slave lasers
is shown in Fig.~\ref{fig:MVAR}.  Below \unit[10]{kHz} acoustic
resonances are visible. At intermediate frequencies $\unit[50]{kHz}
\ldots \unit[1]{MHz}$ distinct narrow peaks corresponding to various
electronic noise sources in the laboratory can be observed. At high
frequencies the \unit[10]{MHz} reference and a $\unit[4.7]{MHz}$
modulation which is used to modulate and lock the master laser can be
identified. Increased noise at $\approx\unit[1]{MHz}$ marks the loop
bandwidth and indicates that the loop gain of at least one laser was
chosen slightly too high.  When sampling the phase difference signal
of the two slaves with a rate of $\unit[5]{MHz}$ we find a root mean
square phase variation of $\unit[0.05]{rad^2}$.

Phase and frequency stability is not the only important
characteristics of a laser. For most of today's high precision
measurements it is also important that the light intensity is stable
to a high degree. Many quantum optics experiments rely on the property
of diode lasers to emit light with intensity noise levels that are
essentially shot noise limited at sideband frequencies over a few tens
of kHz up to intensities of hundreds of microwatt.  Since the phase
lock circuit modulates the diode's injection current there is a
potential risk in producing excess intensity noise when controlling
the laser's phase. However we found the added intensity noise to be
less than $\unit[3]{dB/mW}$ with respect to the shot noise level in a
\unit[500]{kHz} bandwidth. Since the laser diode current modulations
mainly affect the frequency, a phase lock even quietens noisy current
supplies to a certain degree.

\section{Conclusion}
In this paper, we have presented a simple, yet versatile optical phase
lock operating in the frequency range from sub-MHz to 7 GHz.
All-digital phase frequency detection leads to a wide capture range
and renders the circuit robust against amplitude fluctuations of the
beat signal. A special feature of this design is that the reference is
provided by an affordable MHz-range waveform generator. The unit
contains virtually no microwave components, so it can be easily
constructed with little experience in high frequency electronics.

Our circuit has been successfully tested in a number of different
configurations and is an integral component of several experiments
involving electromagnetically induced transparency in atomic
$^{87}$Rb, where the control and signal fields need to be phase locked
to the ground state hyperfine splitting frequency (6.834 GHz)
\cite{figueroa06:_eitdecoherence,vewinger06:_adiab,appel:093602,figueroa08:_compl_charac_squeez_vacuum_propag,macrae08:_DEIT}.
The high stability of the circuit permits long-duration measurements
in atomic coherence experiments. Locking times of many days are
routinely observed.

\bibliography{PLLpaper}

\end{document}